\documentclass[preprint]{aastex}
\newcommand{\be}{\begin{equation}}
\newcommand{\ee}{\end{equation}}
\newcommand{\bea}{\begin{eqnarray}}
\newcommand{\eea}{\end{eqnarray}}

\def\rmin{R_{a,{\rm min}}}
\def\abg{\left (\alpha,\,\beta,\,\gamma\right )}
\def\ec{{\cal E}}
\def\rme{R_\ec}
\def\lc{{\cal L}}
\def\qc{{Q}}
\def\mcal{M}
\def\bec{\bar{\ec}}
\def\be#1{\begin{equation}\label{eq:#1}}
\def\ee{\end{equation}}
\def\EC#1{(\ref{eq:#1})}
\def\bea#1{\begin{eqnarray}\label{eq:#1}}
\def\ee{\end{equation}}
\def\eea{\end{eqnarray}}

\begin{document}

\title{Semi-Analytic Models for Dark Matter Halos}

\author{Lawrence M. Widrow\altaffilmark{1}}
\affil{Department of Physics, Queen's University, 
Kingston, Ontario, Canada K7L 3N6 and\\
Canadian Institute for Theoretical Astrophysics, University of Toronto,
60 St. George St., Toronto M5S 3H8, Canada}

\altaffiltext{1}{widrow@astro.queensu.ca}

\begin{abstract}

Various analytic expressions have been proposed for the density
profile of dark matter halos.  We consider six of these expressions
for which the density profile has a power-law fall-off $\rho\propto
r^{-3}$ at large radii and a power-law cusp $\rho\propto r^{-\gamma}$
($\gamma=0,\,\frac{1}{2},\,1,\,\frac{3}{2}$) at small radii.  The
phase-space distribution function for these models is calculated
assuming spherical symmetry and either an isotropic velocity
dispersion tensor or an anisotropic dispersion tensor of the type
proposed by Osipkov and Merritt.  The differential energy distribution
for these models is also derived.  Several applications are discussed
including the analysis of dark matter search experiments
and the study of halo formation in a cosmological setting.  Analytic
fitting formulae for some of the models are provided.

\end{abstract}


\section{Introduction}

A fundamental problem in modern astrophysics is to determine the
nature of relaxed systems such as galaxies and dark matter halos.
N-body methods are used extensively in this effort and allow one to
follow explicitly the evolution of the phase-space distribution
function (DF) $f$.  A complementary approach seeks to obtain analytic
steady-state models.  Essential to this task is the Jeans theorem
which states that the DF for any equilibrium system can be written in
terms of the integrals of motion.  In addition it is often possible to
make an educated guess as to the form of the DF.  Lynden-Bell (1967),
Tremaine (1987) and Merritt, Tremaine \& Johnstone (1989) for example,
analyze model DFs that are designed to capture the physics of violent
relaxation.  By contrast, Henriksen \& Widrow (1999) propose a DF that
is motivated by the spherical infall model wherein particle energies
vary in a more orderly fashion.

Unfortunately, neither the DFs considered by Merritt, Tremaine, \&
Johnstone (1989) nor the DF considered by Henriksen \& Widrow (1999)
lead to density profiles in agreement with those found in the
simulations.  The alternative is to begin with a desired expression
for the density profile, together with simplying assumptions about the
shape of the velocity ellipsoid, and construct the DF using standard
techniques (e.g., Binney \& Tremaine 1987, hereafter BT).
Occasionally, one is lucky and finds an analytic DF-density profile
pair, as with the Hernquist model (Hernquist 1990).  Though for more
general profiles such as the so-called ``$\gamma$-models'' (see Carollo,
de Zeeuw, and van der Marel (1995) and references therein) numerical
integration is required, the problem is entirely tractible.

The $\gamma$-models, which include the Hernquist model as a special
case, have a density profile with a power-law cusp $\rho\propto
r^{-\gamma}$ at small radii and a $\rho\propto r^{-4}$ fall-off at
large radii.  While these models have proved to be useful in
the study of elliptical galaxies, the dark matter halos found in
cosmological simulations appear to have a power-law fall-off at large
radii that is more gradual than $r^{-4}$.  Navarro, Frenk, \& White
(1996) suggest that the density profiles of dark halos have a
``universal'' shape (the so-called NFW profile) of the form
\be{nfw}
\rho_{\rm NFW}=\frac{\rho_0}{\left (r/a\right )
\left (1+r/a\right )^2}
\ee
They obtain reasonably good fits using this expression for halos that
range in mass from $3\times 10^{11}\,M_\odot$ (dwarf galaxies) to
$3\times 10^{15}\,M_\odot$ (rich galaxy clusters).  However, there is
considerable debate over just what the profile is in the innermost
regions of a halo.  Kravtsov et al.\,(1998) find that the observed
rotation curves of dwarf and low surface brightness galaxies can be
fit by a profile with a shallower central cusp ($\rho\propto
r^{-\gamma}$ where $\gamma\simeq 0.2$) and their own N-body
simulations support this conclusion.  On the other hand Moore et
al.\,(1998) have performed high resolution simulations of
cluster-sized halos and find a central cusp that is steeper than
$r^{-1}$ ($\gamma\simeq 1.5$).

Kravtsov et al.\,(1998) advocate a general fitting formula of the form
\be{general_profile}
\rho=\frac{C\rho_0}{\left (r/a\right )^\gamma
\left (1+\left (r/a\right )^\alpha\right )^{(\beta-\gamma)/\alpha}}
\ee
$\gamma$ controls the slope of the inner profile, $\beta$ that of the
outer profile, and $\alpha$ the sharpness of the transition.  The
normalization parameter $C$ will be discussed below.  The NFW profile
corresponds to $\abg =(1,3,1)$ while the $\gamma$-models correspond to
$\abg=(1,4,\gamma)$ with the Hernquist profile appearing as the
special case, $\gamma=1$.

In this work, we derive semi-analytic DFs for a select subset of the
models described by Eq.\EC{general_profile}.  Specifically, we focus
on models with an $r^{-3}$ density fall-off at large radii
($\beta=3$).  Furthermore, we consider only the six models for which
the gravitational potential can be expressed in terms of elementary
functions, namely $\alpha=1;~\gamma=0,\,\frac{1}{2},\,1,\,\frac{3}{2}$
and $\alpha=2;~\gamma=0,\,1$.

In Section 2 the DFs for the six models assuming an isotropic velocity
dispersion tensor are calculated.  DFs with anisotropic velocity
dispersion of the Osipkov-Merritt type are considered in Section 3 and
their existence and stability is discussed.  In Section 4 the
differential energy distribution various models is calculated.
Several possible applications of our results are discussed in Section
5.  Analytic fitting formulae for many of the DFs found in the text
are provided in the Appendix.

\section{Systems with Isotropic Dispersion Tensors}

For convenience we introduce the dimensionless variables $R\equiv
r/a$, $V\equiv v/\left (4\pi G\rho_0 a^2\right )^{1/2}$,
$\varrho=\rho/\rho_0$ and $F=\left (4\pi G\right
)^{3/2}a^3\rho_0^{1/2}f$.  In addition we define the relative energy
and relative potential (again in dimensionless form) to be
respectively $\ec\equiv -\left (E-\Phi_\infty\right )/4\pi G\rho_0
a^2$ and ${\Psi}\equiv -\left (\Phi-\Phi_\infty\right )/4\pi G\rho_0
a^2$ where $E\equiv\frac{1}{2}v^2+\Phi$ and $\Phi=\Phi(r)$ is the
Newtonian potential with $\Phi(\infty)\equiv\Phi_\infty$.

In general, the Newtonian potential calculated from the density
profile \EC{general_profile} with $\beta=3$ must be determined
numerically.  (For the $\gamma$-models $\left (\alpha=1;~\beta=4\right
)$ an analytic form for the potential not only exists, but can be
inverted to give $r=r(\Phi)$ in closed form.)  However, the potential
can be determined analytically for
$\alpha=1;~\gamma=0,\frac{1}{2},1,\frac{3}{2}$ (Models I-IV) and
$\alpha=2;~\gamma=0,1$ (Models V, VI) and for convenience, we focus on
these six cases.  Expressions for the potentials are collected in
Table 1.  The normalization parameter $C$ in Eq.\EC{general_profile}
is chosen so that for $\Psi(0)=1$ with the limiting form as $R\to 0$
of $\Psi\to 1-Ar^{2-\gamma}$.  Here $A$ is a constant that depends on
$\alpha$ and $\gamma$.  For all models with an $r^{-3}$ power-law
fall-off at large radii, the asymptotic form as $R\to\infty$ is
$\Psi\to A'\ln(R)/R$.


\begin{deluxetable}{ccccc}
\tablecaption{Newtonian Potentials for Selected 
Density Profiles}
\tablewidth{0pt}
\tablehead{\colhead{Model}&
\colhead{$\alpha$}&\colhead{$\gamma$}
&\colhead{$C$}&\colhead{$\Psi$}}
\startdata
I & $1$	   	&   $0$     	& $2$	&
$\frac{2}{R}\ln{(1+R)}-(1+R)^{-1}$\\ 
II & $1$ &   $\frac{1}{2}$     	& $\frac{3}{2}$	&
$1-\frac{3+R}{\left (R+R^2\right )^{1/2}}
+\frac{3}{2R}\ln{S}$\\
III & $1$	   	&   $1$     	& $1$	&
$\frac{1}{R}\ln{(1+R)}$	\\ 
IV & $1$	   	&   $\frac{3}{2}$     	& $\frac{1}{2}$	&
$1-\left (\frac{1+R}{R}\right )^{1/2}
+\frac{1}{2R}\ln{S}$\\ 
V & $2$	   	&   $0$     	& $1$	&
$\frac{1}{R}\sinh^{-1}(R)$	\\ 
VI & $2$	   	&   $1$     	& $\frac{2}{\pi}$&
$~~1-\frac{2}{\pi}\tan^{-1}(R)+
\frac{1}{\pi R}\ln{\left (1+R^2\right )}~~$\\
\enddata
\tablecomments{$S\equiv 1+2R+2\left (R+R^2\right )^{1/2}$}
\end{deluxetable}


The distribution function for an equilibrium spherical system with an
isotropic dispersion tensor depends only on the relative energy $\ec$
and can be calculated from the density profile and potential through
an Abel transform (BT)
\be{abel}
F(\ec)=\frac{1}{\sqrt{8}\pi^2}
\left [
\int_0^{\ec}\frac{d^2\varrho}{d\Psi^2}\frac{d\Psi}{\sqrt{\ec-\Psi}}
+\frac{1}{\sqrt{\ec}}\left (\frac{d\varrho}{d\Psi}\right )_{\Psi=0}
\right ]
\ee
In all of the models that we will consider, the second term on the right
hand side is zero.  

The integral that remains is evaluated numerically.  The integrand
diverges at one or both of the limits but this can be handled
using standard techniques such as those found in Press et al.\,(1986).
The DF is evaluated at values of $\ec$ equally spaced in $\ln{\ec}$
for $0\le\ec < \frac{1}{2}$ and equally spaced in $\ln{(1-\ec)}$ for
$\frac{1}{2}\le\ec\le 1$.  Accuracy is checked by calculating the density 
profile from the DF:
\be{rho}
\varrho(R)=4\sqrt{2}\pi\int_0^\Psi F\left (\ec\right )
\left (\Psi-\ec\right )^{1/2}d\ec
\ee
and comparing with the exact expression.  Typically, $2\times 10^4$
integration points are required to guarantee $0.1\%$ agreement
over the range $R=10^{-6}-10^6$.

In Figure 1, we compare DFs for the Hernquist and NFW profiles.
The DFs in the regime $\left (1-\ec\right )\ll 1$ are nearly identical
(up to a normalization constant).  This is to be expected since in
this regime the systems are dominated by particles at small radii and
the Hernquist and NFW profiles each have an $r^{-1}$ cusp.  While the
DFs diverge in this limit the mass at small radii is finite, as we
will see in Section 5.

The DFs for both the Hernquist and NFW models decrease as $\ec\to 0$.
However, the decrease is slower in the NFW model, a reflection of the
fact that the halo in this model is more extended.  It is
straightforward to determine the functional form of the DF in this
limit.  For the NFW model
\be{asym_1}
\frac{d^2\varrho}{d\Psi^2}\simeq
\frac{1}{R\left (\ln{R}\right )^2}\simeq
\frac{\Psi}{\left (-\ln{\Psi }\right )^3}
\ee
and we find
\be{asym_2}
F(\ec)\propto\frac{\ec^{3/2}}{\left (-\ln{\ec}\right )^3}
\ee
as compared with the Hernquist model where $F\propto\ec^{5/2}$.

The results for our series of $\alpha=1$ models (I-IV) are plotted in
Figure 2.  All of the DFs in this series have a limiting form as
$\ec\to 0$ given, up to a constant, by Eq.\EC{asym_2}.  The difference
in the index for the power-law cusp at small radii is reflected in the
behavior of the DFs as $\ec\to 1$ with the trend that a steeper inner
cusp corresponds to a stronger divergence in this limit.  Once again,
we can determine a limiting form for the DF, this time as
$\ec\to 1$.  For $\gamma=\left (\frac{1}{2},1,\frac{3}{2}\right
)$,~~$d^2\rho/d\Psi^2\propto R^{\gamma-4}$ and we find
\bea{asym_3}
F(\ec)&\propto& \int^\infty_{R_\ec}\frac{dR}
{R^3\left (AR^{2-\gamma}-\left (1-\ec\right )\right )^{1/2}}\\
&\propto &\left (1-\ec\right )^{-(6-\gamma)/(4-2\gamma)}
\eea
where $\rme$ is defined by the relation $\Psi(\rme)=\ec$.  In writing
this expression, we use the fact that the integral is dominated by the
region in $R\simeq\rme\propto \left (1-\ec\right )^{1/(2-\gamma)}$.  
For the NFW profile, $F\propto \left (1-\ec\right
)^{-5/2}$ which is the same as is found in the Hernquist model
(Hernquist 1990).  The case $\gamma=0$ is handled separately: We find
$d^2\rho/d\Psi^2\propto R^{-3}$ and $F\propto \left (1-\ec\right
)^{-1}$.

Kravtsov et al.\,(1998) suggest that the plausible value of the
parameter $\alpha=2$ corresponding to a sharper transition between
inner and outer regions of the halo.  In Figure 3 we compare the DFs
for the two models $\abg = (1,\,3,\,0)$ and $\left
(2,\,3,\,0\right )$.  The most striking difference occurs as $\ec\to
1$ (small radii) where the DF for the $\alpha=2$ model approaches a
constant.  This can be understood by noting that as $R\to 0$,
$d^2\varrho/d\Psi^2$ is finite for $\alpha=2;\,\gamma=0$ but diverges
for $\alpha=1$.

The distinction between the DFs for $\abg=\left (1,\,3,\,1\right )$
and $\left (2,\,3,\,1\right )$ is more subtle.  The asymptotic forms
as $\ec\to 1$ and $\ec\to 0$ are the same and so the difference between
the models arises solely in the transition region.  In Figure 4 we plot
the ratio of the DFs for the two models.  If we normalize the models
at $\ec=1$ ($R=0$) then as $R\to\infty$ the DF for $\alpha=2$ will 
exceed that for $\alpha=1$ by a factor $\simeq 6$.

Physical DFs ($F\ge 0$ for $0\le\ec\le 1$) exist for all of the
isotropic models considered here.  Moreover, both $F$ and
$d^2\varrho/d\Psi^2$ are monotonically increasing functions of $\ec$.
This, by Antonov's second and third laws, is sufficient to guarantee
that these models are stable against both radial and nonradial
perturbations (BT).

\section{DFs with Anisotropic Velocity Dispersion} 

The DF of any steady-state system that is spherically symmetric in
both velocity and configuration space can be expressed as a function
$f(E,L)$ where $L$ is the magnitude of the angular momentum vector
(BT).  In general, the velocity-dispersion tensors for these models
are anisotropic: the velocity dispersions in the two tangential
directions are equal but in general different from the velocity
dispersion in the radial direction.  We investigate a special class of
these models in which the DF assumes the form (Osipkov 1979; Merritt
1985a,b)
\be{om}
F(\ec,\lc)=F(\qc)
\ee
where $\lc=L/\left (4\pi G\rho_0 a^4\right )^{1/2}$ and $\qc\equiv\ec
- \lc^2/2R_a^2$.  In addition, the condition $F=0$ for $\qc\le 0$ is
imposed.  $R_a$ is often called the anisotropy radius: Outside $R_a$
the velocity dispersion is peaked toward radial orbits while inside
$R_a$ the dispersion is nearly isotropic.  As discussed in Carollo, de
Zeeuw, and van der Marel (1995), physical models do not exist for
$R_a<\rmin$ where $\rmin$ depends on the shape of the potential and
must be determined numerically.  For values of $R_a$ below this
threshold, the DF must become negative for a range of values in $Q$ in
order to compensate for the excess population of radial orbits needed
to produce the halo at large radii.  The problem is more severe for
the systems considered here since the density profile at large radii
varies as $r^{-3}$ rather than $r^{-4}$.  In particular, we find that
$\rmin\simeq 0.36$ and $0.75$ for $\abg=\left (1,\,3,\,0\right)$ and
$\left (1,\,3,\,0\right)$ respectively.  By comparison, Carollo, de
Zeeuw, and van der Marel (1995) find $\rmin\simeq 0.20$ and $0.45$ for
$\abg=\left (1,\,4,\,0\right)$ and $\left (1,\,4,\,0\right)$
respectively (cf. their Figure 1).

The DF for the Osipkov-Merritt models is found by replacing the
$\varrho$ in Eq.\EC{abel} with the auxiliary density $\varrho_Q\equiv
\left (1+R^2/R_a^2\right )\varrho$.  The results for a sequence of
Osipkov-Merritt NFW models is shown in Figure 5.  The pathological
nature of the DF as $R_a\to\rmin$ is evident in the dotted curve
($R_a=\rmin+\epsilon$ where $\epsilon\sim 10^{-5}$).  

Notice that for these models the DF at small $\ec$ is a decreasing
function of $\ec$ ($F\propto\ec^{-1/2}$).  Stability analysis against
radial perturbations by application of Antonov's second law is
therefore inconclusive (BT).  While models with $R_a\to \infty$ are
stable against both radial and non-radial perturbations those with
$R_a\to 0$ are almost certainly unstable to radial perturbations
(Merritt 1985b).  Numerical experiments are therefore required to
determine the exact region of stability.

\section{Differential Energy Distribution}

When comparing results of N-body simulations with those from analytic
models, it is natural to use the differential energy distribution
$dM/d\ec$ (BT).  For models with an isotropic
velocity dispersion tensor, $d\mcal/d\ec$ is simply the product of the
DF with the density of states: 
\be{dmde}
\frac{d\mcal}{d\ec}=F(\ec)G(\ec) \ee where 
\be{dofs}
G(\ec)=16\sqrt{2}\pi^2\int_0^{\rme}\left (\Psi-\ec\right )^{1/2} R^2dR
\ee 
$\mcal$ is a dimensionless mass variable scaled by $\rho_0 a^3$.  Note
that as $\ec\to 0$, $G(\ec)\propto\ec^{-5/2}$ (up to logarithmic
corrections) for both the models considered here and the Hernquist
model.

Figure 6 is a plot of $d\mcal/d\ec$ for models I-V and for the
Hernquist model.  In the limit $\ec\to 0$, $d\mcal/d\ec$ for the
Hernquist model approaches a constant reflecting the fact that the
total mass is finite.  Conversely, for the models considered here,
$d\mcal/d\ec$ diverges as $\ec^{-1}$, a symptom of the logarithmic mass
divergence at large radii.

It is a somewhat more difficult exercise to calculate the differential
energy distribution for Osipkov-Merritt models.  Let us begin, as
is done in the derivation of 
Eq.\EC{dmde}, with the expression for the total mass
within a radius $R$ (BT):
\be{tm1}
\mcal(R)=16\pi^2\int_0^R R'^2 dR'\int_0^{\Psi(R')} v d\ec
\int_0^{\pi/2}F(\ec,RV\sin{\eta})\sin{\eta}d\eta
\ee
where $\eta$ is the polar angle in velocity space: 
$V_r\equiv V\cos{\eta}$.  We first replace the integration variable
$\eta$ with $Q$ using the relation $\qc\equiv \ec-\frac{1}{2}\left (
R/R_a\right )^2V^2\sin^2{\eta}$ 
and then interchange the $R$ and $\ec$ integrations to find
\be{tm2}
\mcal(R)=8\sqrt{2}\pi^2 R_a\int_0^1 d\ec\int_{0}^{\rme} 
R'dR'\int_{\ec-\bar{\ec}}^{\ec}
\frac{F(\qc)d\qc}
{\left (\qc-\left (\ec-\bec\right )\right )^{1/2}}
\ee
where $\bec\equiv\frac{1}{2}\left (R'/R_a\right )^2V^2$.
Thus, the differential energy distribution for the Osipkov-Merritt
models can be evaluated by performing the double integral
\be{dmdeOM}
\frac{d\mcal}{d\ec}=
8\sqrt{2}\pi^2 R_a\int_{0}^{R_{\rm min}}R' 
dR'\int_{\ec-\bar{\ec}}^{\ec}\frac{F(\qc)d\qc}
{\left (\qc-\left (\ec-\bec\right )\right )^{1/2}}
\ee
where $R_{\rm min}={\rm min}\left (\rme,\,R\right )$.
It is straightforward to show that in the limit $R_a\to\infty$,
$\bec\to 0$ and Eq.\EC{dmdeOM} reduces to \EC{dmde}.

The results for the four NFW models considered in Section 3 are shown
in Figure 7.  Note that the divergence as $\ec\to 0$ is even more
severe ($dM/d\ec\propto\ec^{-2}$).

\section{Applications}

In this section three possible applications of the results presented
above are briefly discussed.

\subsection{Initial Conditions for N-body Experiments}

The dynamic range in N-body simulations has now reached the level
where it is possible to study individual halos in exquisite detail
while maintaining an accurate representation of the large-scale tidal
fields that shape them.  Nevertheless, it is often useful to conduct
highly controlled, albeit artificial, numerical experiments in order
to gain a better understanding of the processes at work during halo
formation.  For example, numerous groups have used numerical
simulations to study the accretion of satellites with a parent halo as
well as the merger of two halos of comparable mass.  Another set of
experiments follows the evolution of a gas and dark matter system
inorder to better understand the process of disk formation.
Typically, very simple DFs (e.g., truncated isothermal sphere, Plummer
model) are used to set up the initial conditions for these experiments
and the choice is often made out of convenience, i.e., the
availability of analytic DFs, rather than an expectation of what might
arise in a cosmological setting.

The DFs found in this work allow one to set up a variety of halo
models with various forms for the density profile and velocity 
distribution tensor.  For the models
with isotropic velocity dispersion tensors, the probability for a 
particle to have energy $\ec$ and radius $R$ is
\be{probability}
P\left (\ec,\,R\right )~\propto~
R^2\left (\Psi-\ec\right )^{1/2}F\left (\ec\right )
\ee
The rejection method described, for example, in Press et al.\,(1986),
provides a simple technique for selecting particles from this
distribution.  Once $\ec$ and $R$ for a given particle are known, the
speed $V$ can be determined immediately.  One then chooses, at random,
four angles (two in configuration space and two in velocity space) to
yield the six phase space coordinates of the particle.  This procedure
can be extended easily to the Osipkov-Merritt models.

\subsection{Interpretation of Results from Dark Matter Search Experiments}

The announcements in 1993 by the MACHO (Alcock et al.\,1993) and EROS
(Aubourg et al.\,1993) collaborations of candidate microlensing events
toward the LMC have highlighted the need for self-consistent model DFs
of the Galaxy's halo.  Interpreting the results from these experiments
requires a comparison of the observed and predicted event rates.  The
latter has now been calculated for a set of halo models too numerous
to list here.  These models are generally constructed in one of two
ways.  One can begin with an ansatz for the mass distribution (e.g.,
triaxial spheroid with a prescribed density law) and assume a simple
form for the velocity dispersion tensor (e.g., isotropic and
Maxwellian).  Of course, models constructed in this manner do not, in
general, correspond to true equilibrium systems, i.e., solutions of
the time-independent collisionless Boltzmann equation.  A second
approach employs exact, analytic model DFs.  Here, the Jeans theorem
is invaluable since it allows one to construct model DFs simply by
taking functions of the integrals of motion.  The so-called power-law
models in which the DF is constructed from powers of the energy and
the angular momentum vector $L_z$ (Evans 1993) have been used in this
way (Evans \& Jijina 1994).

The DFs presented in this work provide an alternative set of models
suitable for the analysis and interpretation of results from dark
matter search experiments.  The density profiles include many of the
popular forms found in the literature.  In addition, the
Osipkov-Merritt ansatz for velocity space anisotropy is consistent
with what one expects for a realistic halo model, i.e., primarily
radial orbits in the outermost regions of the halo where particles
have only recently separated from the Hubble flow.

\subsection{Formation of Dark Matter Halos}

The hierarchical clustering hypothesis provides a compelling picture
for the formation of structure in a Universe dominated by cold,
dissipationless dark matter.  This scenario naturally lends itself to
the development of phenomenological models for the growth of dark
matter halos (e.g., the Press-Schechter formalism (Press and Schecter
1974)).  Nevertheless, hierarchical clustering says little about the
internal structure of these systems.  For this, one must understand
the process by which a collisionless system relaxes to an equilibrium
or quasi-equilibrium state.  To this end, two opposing pictures have
emerged.  The first suggests that violent relaxation is the dominant
process at work, the conjecture being that particles will transfer
energy to one another as they move through the rapidly varying
potential of the collapsing system.  In so doing, the particles lose
all memory of their initial state.  The alternative picture is based
on the spherical infall model (Gunn and Gott 1972; Fillmore and
Goldreich 1984; Bertschinger 1985) in which a system relaxes from the
inside out.  Particles near the peak of the initial density
perturbation collapse first and constitute the most tightly bound
regions of the final system.  Likewise, particles in the wings of the
initial density perturbation collapse later on and form the system's
outer halo.  Here, one expects a direct correspondence between initial
conditions and the final state of the system and in particular, a
tight correlation between initial and final energy and angular
momentum.

Not surprisingly, the dark matter halos found in cosmological
simulations appear to follow an intermediate path in reaching a
relaxed or virialized state.  Quinn and Zurek (1988) track the binding
energy and angular momentum of selected particles during the simulated
formation of a dark matter halo.  They conclude that though energy and
angular momentum are not conserved, the collapse is more orderly than
one would predict assuming violent relaxation.  Indeed there seems to
be a tight correlation between initial and final energy and angular
momentum.  The suggestion then is that there is a simple relation
between the differential energy and angular momentum distributions of
a relaxed system and the initial perturbation that gave rise to it.
The DFs presented in this work may provide a further link between the
$dM/d\ec$ or $d^2M/d\ec dL$ and the density profile.

\section{Conclusion}

The density profile and associated gravitational potential provide a
popular way of characterizing dark matter halos.  To be sure, the
halos found in simulations are varied in shape and rich in
substructure.  However, it is essentially the density profile that
determines the contribution by the halo to the observed rotation
curve.  Moreover, the evolution of the gas component (e.g., in forming
a disk galaxy) depends sensitively on the shape and depth of the
potential.

In this paper, we have presented DFs that reproduce six different
density profiles.  All of the models assume spherical symmetry and a
power-law fall-off at large radii $\rho\propto r^{-3}$ but differ in
the slope of the inner power-law cusp and the sharpness of the
transition in going from inner to outer regions of the halo.

We have considered models with velocity-space anisotropy of the type
proposed by Osipkov and Merritt.  The key feature and also limitation
of these models is that the distribution of orbits varies from one
that is nearly isotropic in the inner regions of the halo to one that
is nearly radial in the outer regions.  In addition, since the
Osipkov-Merritt models provide a family of DFs, parametrized by $R_a$,
that reproduce the same density profile, they may be combined to yield
a fairly general class of models (Merritt 1985a).

Though the DFs presented in this work were derived numerically the
analytic fitting formulae provided in the Appendix should enable
researchers to use them with the same ease as they would find with
closed form DFs such as the Hernquist model.

\acknowledgments{I would like to thank the Canadian Institute 
for Theoretical Astrophysics for their hospitality during a sabbatical 
visit.  I also acknowledge R. Henriksen for useful discussions.
This work is supported by the Natural Sciences and Engineering Research
Council of Canada}

\newpage

\appendix

\section{Appendix}

In this appendix, we provide analytic fitting formulae for some of the
DFs presented in the text.  The limiting forms for the isotropic models,
as discussed in the text, are:
\be{limiting_forms}
F(\ec)~\propto\left\{
	\begin{array}{ll}
	\ec^{3/2}\left (-\ln{\ec}\right )^{-3} 
		& \mbox{for $\ec\to 0$}\\
	\left (1-\ec\right )^{-\lambda}
		& \mbox{for $\ec\to 1$}
	\end{array}
\right.
\ee
where $\lambda=\left (1,\,\frac{11}{6},\,\frac{5}{2},\,\frac{9}{2}
,\,0,\,\frac{5}{2}\right )$ for Models I-VI respectively.
Noting that $-\ln{\ec}\simeq \left (1-\ec\right )$ for $1-\ec\ll 1$,
we propose the following fitting formulae for the isotropic DFs:
\be{fitting}
F(\ec)=F_0\ec^{3/2}\left (1-\ec\right )^{-\lambda} 
\left (\frac{\left (-\ln{\ec}\right )}{1-\ec}\right )^{q}e^{P}
\ee
where the polynomial $P\equiv \sum_i p_i\ec^i$
is introduced to improve the fit.  In general, 3-5 terms
in $P$ are required to achieve a reasonable fit.  The parameters
are given in Table 2.  Notice that $q\simeq -3$ as is expected
from Eq.\EC{limiting_forms}.


\begin{deluxetable}{ccccccccc}
\tablecaption{Fitting Formula Parameters for Isotropic Models}
\tablewidth{0pt}
\tablehead{
\colhead{Model}&\colhead{$F_0$}&\colhead{$q$}&\colhead{$p_1$}&
\colhead{$p_2$}&\colhead{$p_3$}&\colhead{$p_4$}&\colhead{$p_5$}&
\colhead{$p_6$}}
\startdata
I  &$2.0460\times 10^{-2}$&-2.7145&-1.0215
	&23.766&-98.330&194.50&-180.01&63.296\\
II &$3.6478\times 10^{-2}$&-2.7092&0.8670
	&-10.035&65.895&-166.31&179.21&-70.007\\
III&$9.1968\times 10^{-2}$&-2.7419&0.3620&-0.5639&-0.0859&-0.4912&&\\
IV &$4.8598\times 10^{-1}$&-2.8216&0.3526&-5.1990&3.5461&-0.8840&&\\
V  &$5.8807\times 10^{-2}$&-2.6312&-3.7147
	&41.045&-132.20&216.90&-170.23&51.606 \\
VI  &$1.4696\times 10^{-1}$&-2.6210&-3.6125
	&23.172&-78.104&135.80&-123.11&43.705\\
\enddata
\end{deluxetable}


For the anisotropic models discussed in the text
($\alpha=1;\,\gamma=1$), we use the fitting formula
\be{fitting2}
F(Q)=F_0 Q^{-1/2}\left (1-Q\right )^{-\lambda} 
\left (\frac{\left (-\ln{Q}\right )}{1-Q}\right )^{q}e^{P}
\ee
where $P$ is now a polynomial in $Q$.  The parameters are given
in Table 3.  No attempt is made to fit the DF for 
$R_a\simeq\rmin$.


\begin{deluxetable}{ccccccccc}
\tablecaption{Fitting Formula Parameters for Anisotropic Models}
\tablewidth{0pt}
\tablehead{
\colhead{$R_a$}&\colhead{$F_0$}&\colhead{$q$}&\colhead{$p_1$}&
\colhead{$p_2$}&\colhead{$p_3$}&\colhead{$p_4$}&\colhead{$p_5$}&
\colhead{$p_6$}}
\startdata
0.6 &$1.0885\times 10^{-1}$&-1.0468&-1.6805
	&18.360&-151.72&336.71&-288.09&85.472\\
1&$3.8287\times 10^{-2}$&-1.0389&0.3497&-12.253&-9.1225&101.15
	&-127.43&47.401\\
3 &$4.2486\times 10^{-3}$&-1.0385&0.7577&-25.283&149.27&-282.53
	&229.13&-69.048\\
10  &$3.8951\times 10^{-4}$&-1.0447&-2.2679
	&79.474&-237.74&329.07&-223.43&59.581 \\
\enddata
\end{deluxetable}


{}

\begin{figure}
\caption{DFs as a function of relative energy $\ec$
for the NFW profile (solid curve) and Hernquist 
profile (dashed curve).}
\plotone{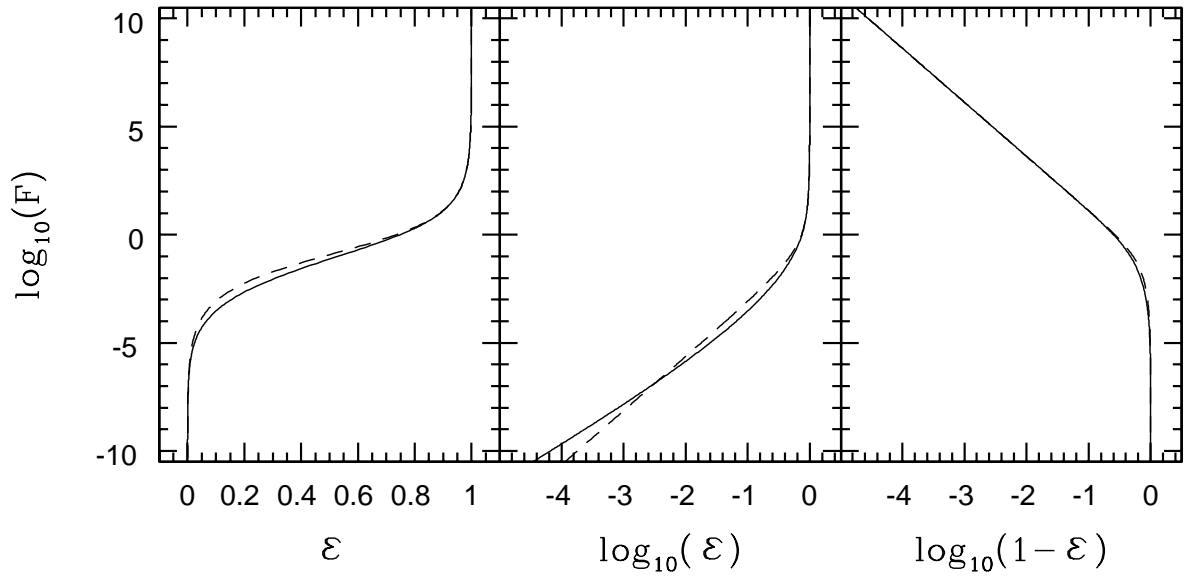}
\end{figure}

\begin{figure}
\caption{DFs for Models I-IV ($\alpha=1;\,\beta=3$):
Model I, $\gamma=0$ (dotted curve); Model II, $\gamma=\frac{1}{2}$
(long-dashed curve); Model III, $\gamma=1$, the NFW profile (solid
curve); Model IV, $\gamma=\frac{3}{2}$ (dot-dashed curve).}
\plotone{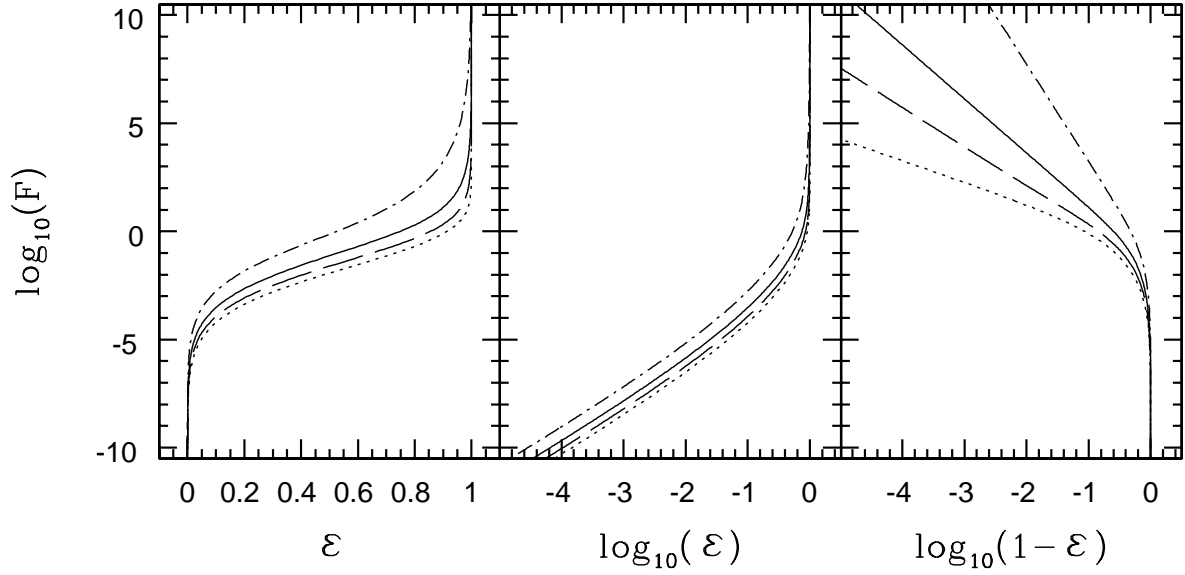}
\end{figure}

\begin{figure}
\caption{Comparison of the DFs for Models I and V
($\gamma=0;\,\beta=3$): $\alpha=1$ (dotted curve) and $\alpha=2$
(dot-long dashed curve).}
\plotone{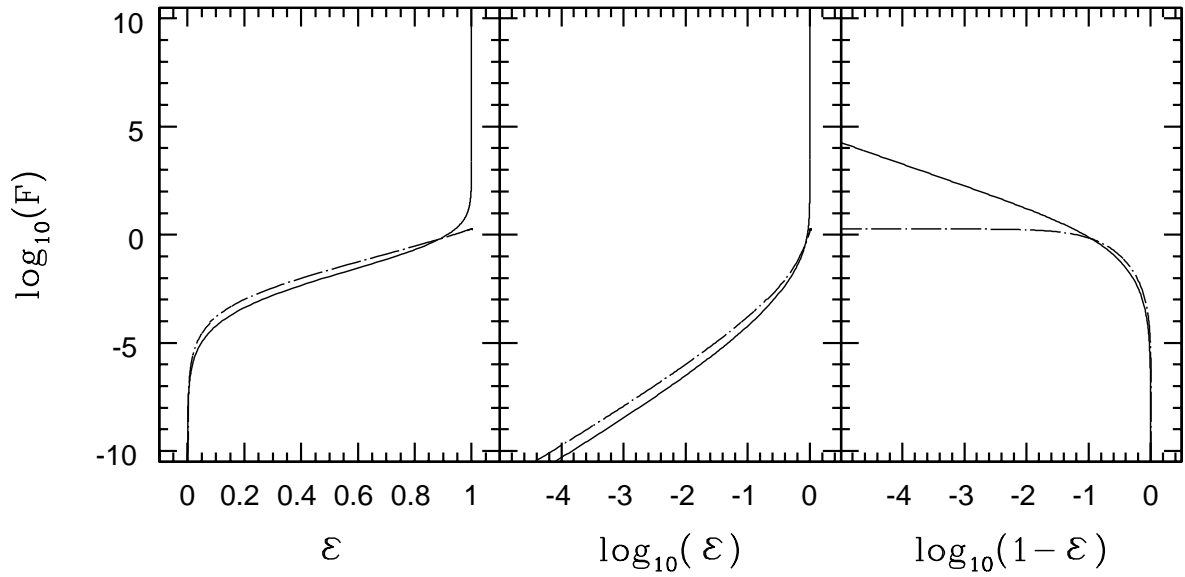}
\end{figure}

\begin{figure}
\figcaption[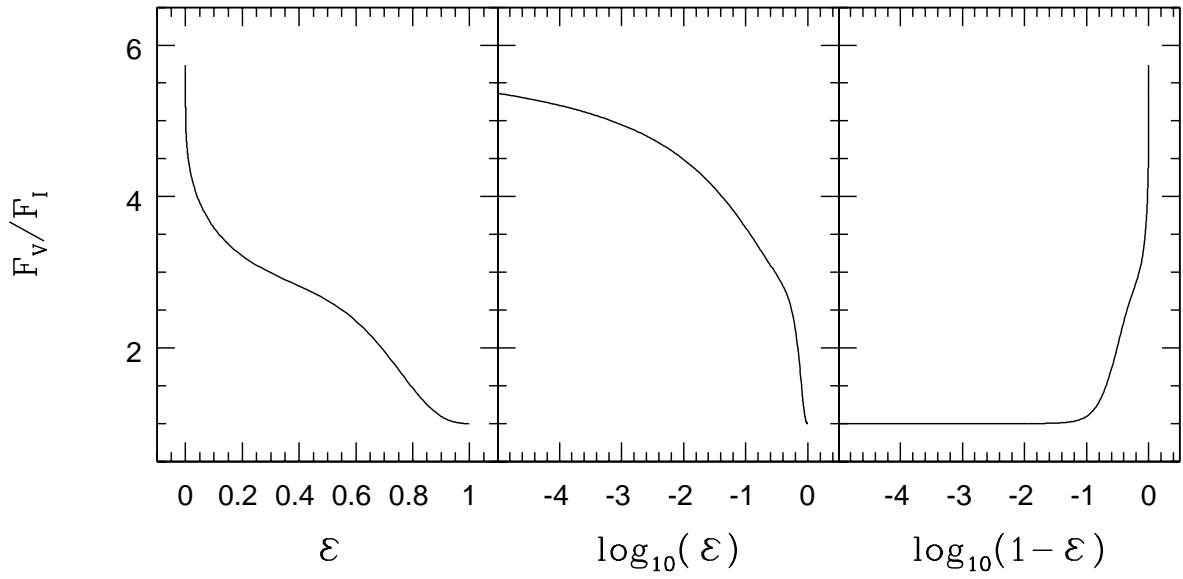]{DF for Model VI divided by the DF for Model II}
\plotone{fig4.ps}
\end{figure}

\begin{figure}
\caption{DF for the NFW profile assuming a velocity
dispersion tensor of the type proposed by Osipkov and Merritt.  The
different curves correspond to different values of the anisotropy
parameter $R_a$: $R_a=0.35548$ (dotted curve); $R_a=0.6$ (dashed
curve);$R_a=1$ (long-dashed curve); $R_a=3$ (dot-dashed curve);
$R_a=10$ (dot-long dashed curve).  The DF assuming an isotropic
dispersion tensor $R_a\to\infty$ (Model III) is given by the solid
curve.}
\plotone{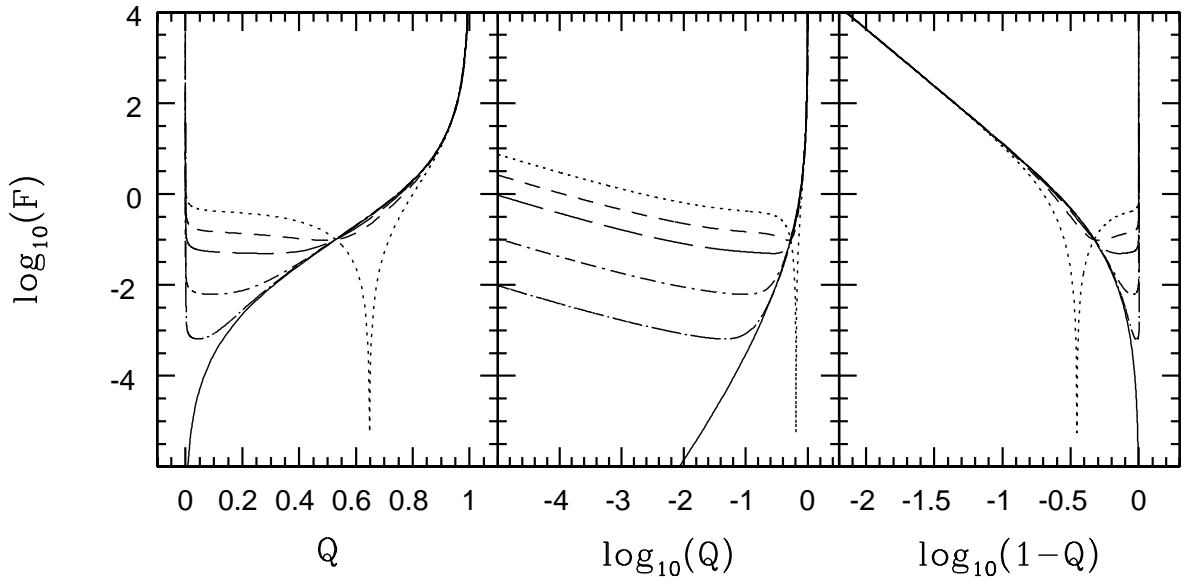}
\end{figure}

\begin{figure}
\figcaption{Differential energy distribution, $dM/d\ec$
as a function of energy for Models I-V and the Hernquist model.
Curves are labeled as in Figures 1 and 2}
\plotone{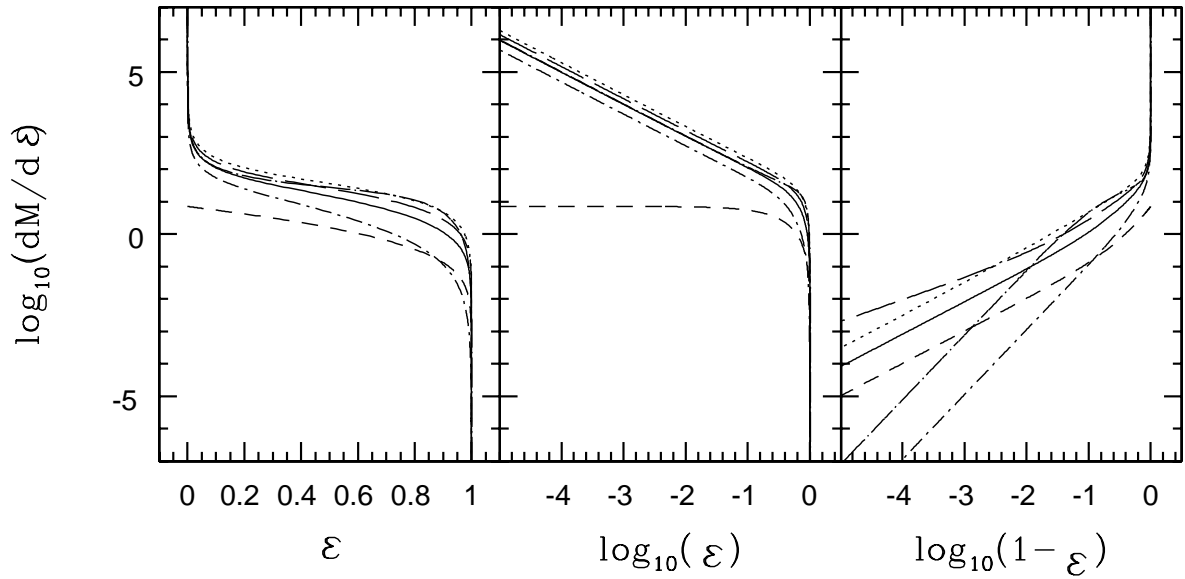}
\end{figure}

\begin{figure}
\figcaption{$dM/d\ec$ for the anisotropic models.  Curves
are labeled as in Figure 5}
\plotone{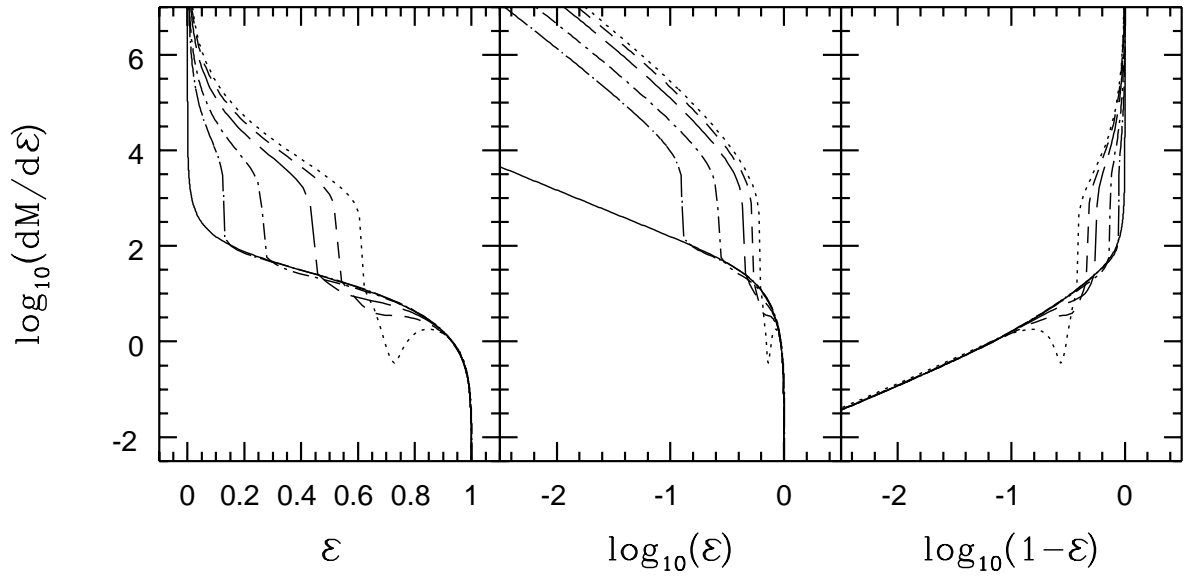}
\end{figure}

\end{document}